\documentclass[modern]{aastex631}
\usepackage{bbm}
\shorttitle{Transit Detection}
\shortauthors{Cui et al.}
\graphicspath{{./}{figures/}}

\begin{document}
\title{Identify Light-Curve Signals with Deep Learning Based Object Detection Algorithm. I. Transit Detection}

\correspondingauthor{Kaiming Cui}
\email{cuikaiming@sjtu.edu.cn}

\author[0000-0003-1535-5587]{Kaiming Cui}
\affiliation{Tsung-Dao Lee Institute, Shanghai Jiao Tong University, 800 Dongchuan Road, Shanghai 200240, People’s Republic of China}
\affiliation{CAS Key Laboratory of Optical Astronomy, National Astronomical Observatories, Chinese Academy of Sciences, Beijing 100101, China}
\affiliation{School of Astronomy and Space Sciences, University of Chinese Academy of Sciences, Beijing 100049, China}

\author{Junjie Liu}
\affiliation{Megvii Research Insitute, Megvii Technology, Chengdu, Sichuan 610095, China}

\author[0000-0001-6039-0555]{Fabo Feng}
\affiliation{Tsung-Dao Lee Institute, Shanghai Jiao Tong University, 800 Dongchuan Road, Shanghai 200240, People’s Republic of China}

\author{Jifeng Liu}
\affiliation{CAS Key Laboratory of Optical Astronomy, National Astronomical Observatories, Chinese Academy of Sciences, Beijing 100101, China}
\affiliation{School of Astronomy and Space Sciences, University of Chinese Academy of Sciences, Beijing 100049, China}
\affiliation{WHU-NAOC Joint Center for Astronomy, Wuhan University, Wuhan, Hubei 430072, China}



\begin{abstract}
Deep learning techniques have been well explored in the transiting exoplanet field; however, previous work mainly focuses on classification and inspection. In this work, we develop a novel detection algorithm based on a well-proven object detection framework in the computer vision field. Through training the network on the light curves of the confirmed Kepler exoplanets, our model yields about 90\% precision and recall for identifying transits with signal-to-noise ratio higher than 6 (set the confidence threshold to 0.6). Giving a slightly lower confidence threshold, recall can reach higher than 95\%. We also transfer the trained model to the TESS data and obtain similar performance. The results of our algorithm match the intuition of the human visual perception and make it useful to find single-transiting candidates. Moreover, the parameters of the output bounding boxes can also help to find multiplanet systems. Our network and detection functions are implemented in the {\tt Deep-Transit} toolkit, which is an open-source Python package hosted on Github and PyPI.

\end{abstract}



\section{Introduction} \label{sec:intro}
The increased amount of data in astronomy makes it necessary to apply machine-learning algorithms for regression, classification, detection, clustering, forecasting, etc. As a typical branch of machine-learning methods, deep learning algorithms have been rapidly evolving in recent years, possessing better performance compared to traditional methods in image classification \citep{krizhevsky2012imagenet}, visual recognition \citep{Girshick2014}, and many other areas. Some featured algorithms, such as convolution neural network \citep[CNN;][]{Lecun1998}, recurrent neural network \citep[RNN;][]{rumelhart1985learning}, long short-term memory \citep[LSTM;][]{Hochreiter1997}, and generative adversarial networks \citep[GAN;][]{2014Generative}, are also used in astronomy. For example, \citet{Leung2019MNRAS.483.3255L} develop the {\tt astroNN} package for spectra analysis, and \citet{Charnock2017ApJ...837L..28C} and \citet{Muthukrishna2019PASP..131k8002M} use RNN for transients classification on multiband photometric time series. \citet{Liu2019ApJ...877..121L} use LSTM to predict solar flares based on some active region information and flare histories.
\citet{Schawinski2017MNRAS.467L.110S} use GAN for recovering galaxy morphology.

In the field of signal detection and classification of light curve, 1D CNN, RNN, and LSTM are commonly used \citep[e.g.,][]{Hinners2018AJ....156....7H, Feinstein2020AJ....160..219F}. As a very important and attractive project, finding transiting exoplanets from light curves also makes extensive use of deep learning techniques, with 1D CNN being the primary method.
\citet{Pearson2018MNRAS.474..478P} first use 1D CNN to search for exoplanets based on iterative use of a binary classifier, and \citet{Zucker2018AJ....155..147Z} independently develop a similar work with different data set. Meanwhile, a series of work on exoplanet vetting based on 1D CNN starting from \citet{Shallue2018AJ....155...94S}, who work on Kepler data with local and global views for folded light curves of transit candidates. Following works introduce external information \citep[e.g.,][includes centroid curves and stellar parameters]{Ansdell2018ApJ...869L...7A} or applied it to K2 data set \citep{Dattilo2019AJ....157..169D}, TESS data set \citep{Yu2019AJ....158...25Y, Osborn2020A&A...633A..53O}, Wide Angle Search for Planets program \citep{Schanche2019MNRAS.483.5534S}, and the Next Generation Transit Survey \citep{Chaushev2019MNRAS.488.5232C}.
\citet{Olmschenk2021AJ....161..273O} develop a systematic pipeline for detecting and vetting transiting exoplanets from TESS full-frame image (FFI) light curves. They apply a 1D CNN for outputting the confidence of existing potential transiting exoplanets in a given FFI light curve.

In this work, we explore a two-dimensional (2D) object detection algorithm to identify transiting signals. 2D means the detection is performed on the image. Compared with previously 1D CNN, our approach is straightforward and matches human visual intuition. We implement our neural network based on Kepler, then transfer learning to TESS. Section \ref{sec:training} introduces the Kepler training data preparation (Section \ref{subsec:data-prep}), network architecture (Section \ref{subsec:network}), detailed training steps (Section \ref{subsec:training}), and test and comparison results (Section \ref{subsec:kepler-test}). In Section \ref{sec:transfer}, we transfer the model trained on Kepler to TESS data. After we have a trained model, further applications and inspections can be developed based on our results. Section \ref{sec:application} shows several application examples, and Section \ref{sec:discuss} discusses features and limitations of our method. Our training and detection algorithms are implemented in the {\tt Deep-Transit}, which is an open-source Python package. \footnote{\url{https://github.com/ckm3/Deep-Transit}}

\section{Implementation and Training} \label{sec:training}
Different from the commonly used 1D CNN and RNN (LSTM) algorithm, we plot the lightcurve to an image, so that the 1D time series is converted into a 2D image. This allows the rapidly developing computer vision algorithms to be applied to light-curve data. In terms of data preparation and network structure, our approach has two main advantages over previous algorithms.

\begin{enumerate}
    \item The sampling frequency requirement for light curves is relatively lenient. For 1D CNN and RNN (LSTM), the input dimension is a fixed number of data points, so that different data set require interpolation or binning. In contrast, our method is less sensitive to the sampling frequency and preserves the original information of the data.
    
    \item Our network has an output combining three scales: large, medium, and small. Thus it can be adapted to different information densities generated by different data sets without modifying network structure.
\end{enumerate}

\subsection{Data Set Preparation}\label{subsec:data-prep}
As one of the most successful light-curve products, Kepler light curves are ideal for our work. 
Our data sets are selected from the Kepler light curves for confirmed transiting planets \footnote{\url{https://exoplanetarchive.ipac.caltech.edu/docs/counts_detail.html}}. The time, durations, and periods of those transits are collected from the Kepler Threshold-Crossing Event catalog \citep[][]{Twicken2016AJ....152..158T}. Compared with synthetic data sets, real data have more realistic stellar activity signal and systematics. Before creating our training and validation data, we randomly select the Kepler light curves for 100 confirmed exoplanets as the test data.

Similar to other transit detection algorithms, we also need to detrend instrumental and the photometric variation caused by stellar activity. To do so, we apply the Tukey’s biweight algorithm \citep[][]{mosteller1977data} implemented by Wotan \footnote{\url{https://github.com/hippke/wotan}}, whose performance is proven in blind search of Kepler and K2 data \citep{Hippke2019A&A...623A..39H}. The parameters we use are the default parameters in Wōtan. Then, we have a sigma clipping for the detrended light curves. The upper clipping limit is three standard deviations and the lower clipping limit is 20 standard deviations.

Since our method relies on the visual presentation of the data, the method of conversion from light-curve data to images is critical.
In our work, we first split the light curve into 30 day segments (less than 30 days are not split) and make the next window overlap the previous window by 5 days (i.e., a 30 day sliding window size with a 25 day step). Then, in each 30 day segment, the light curve is split into 10 day segments, and the next segment overlaps the previous one by 3 days (i.e., a 10 day sliding window with a 7 day step). 
We generate images for each of the 10 day light-curve segments. 
We choose the window length based on our experience and data propriety of Kepler and TESS, and it can be modified for different type of signals. Actually, after training, a slight modification of the window length has little effect on the results.
For convenience, our image size is chosen to be $416 \times 416$, which is a commonly used size for object detection. Larger size could has better performance but cost more computation time.

Based on our experience in light-curve identification, we plot data in line with point markers to enhance the characteristics of the light curve. The line width is set to 1 pixel, and the data point is the circle with a 2-pixel radius. 
These choices are flexible, because after training is complete, the network has the ability to generalize and can handle slightly different plotting styles.
To keep things as simple as possible, all the lines and points are black, so the image is a single-channel bitmap with the shape of (416, 416, 1). The flux range for plotting is the flux range of 30 day segments, rather than within 10 days. This is because we want our model to have some perception of a wider range of time and avoid some local false positives. Also, we extend the lower limit of 30 day segments' flux by 2\% to avoid the lowest flux of a transit being at the edge of an image. 

\begin{figure}
    \centering
    \plotone{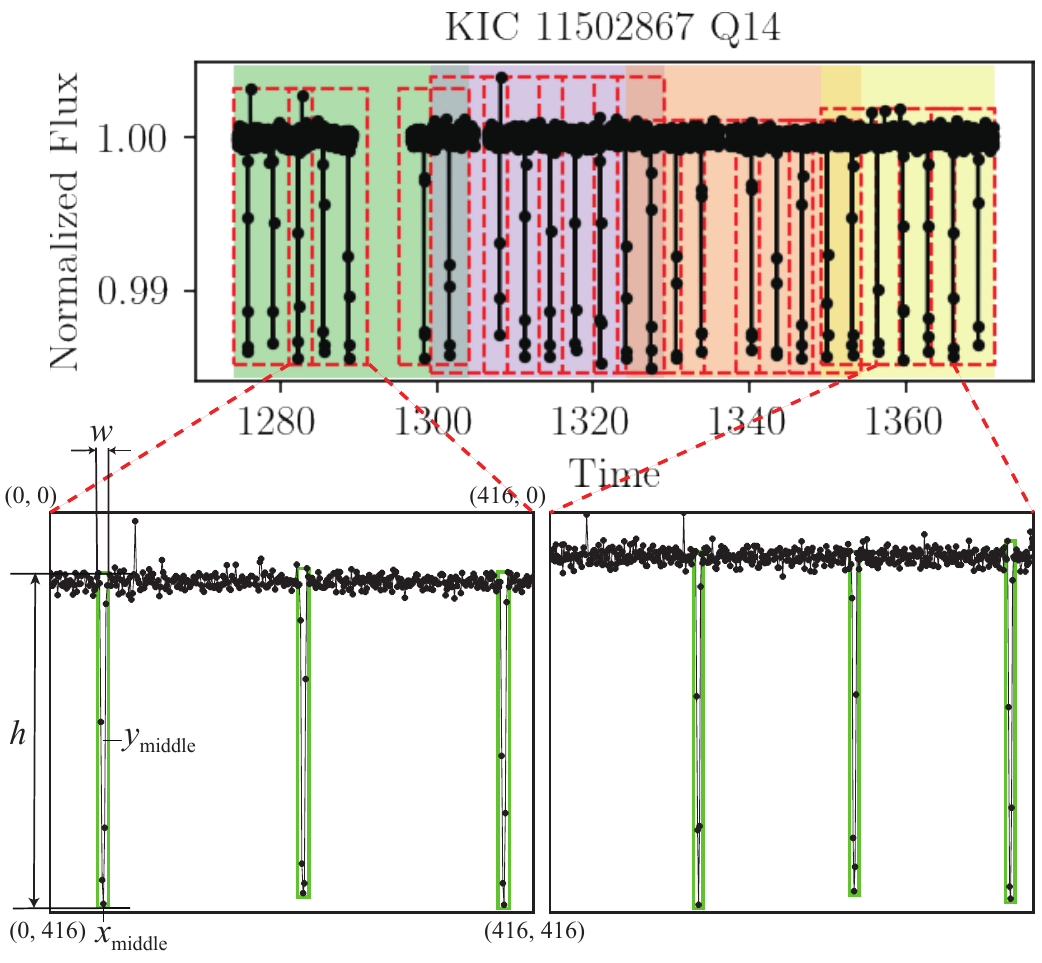}
    \caption{Schematic diagram of training data set generation steps. The top panel shows a part of a detrended light curve. The color shaded sections indicate 30 day windows, and they have 5 day overlapped areas between adjacent windows. The dashed red lines are 10 day selection ranges, and they have 3 day overlapped areas between adjacent ranges. The bottom two panels are the zoom-ins of two 10 day selection ranges, and the bounding boxes are also plotted with green rectangular boxes. The coordinates of a bounding box $(x_{\mathrm{middle}}, y_{\mathrm{middle}}, w, h)$ and four corners are indicated in the bottom left panel.}
    \label{fig:data-pre}
\end{figure}

The format of our bounding box label is $(x_{\mathrm{middle}}, y_{\mathrm{middle}}, w, h, \mathrm{S/N})$. The $x_{\mathrm{middle}}$ and $y_{\mathrm{middle}}$ are the normalized coordinates of the center of a transit (i.e., convert the coordinates to 0--1, the coordinate of the left edge is 0, the right edge is 1, the top edge is 0, and the bottom edge is 1). The width $w$ of our bounding box is set to twice the normalized duration, and if the width is less than 0.02, then enforce it to 0.02. The top value of the bounding box is 95\% of the flux of the 10 day segment. To ensure that the bounding box can cover the whole transiting region, we extend the height $h$ down by 4 pixels. We also calculate the signal-to-noise ratio (S/N) for each transit. The median and standard deviation are derived from the out-of-transit region of the 10 day segment. The depth of transit is estimated from the median value of the middle half. In our training set, only transits with S/N $>$ 3 are generated. Figure \ref{fig:data-pre} is a schematic diagram of our training data set generation.

After the data set preparation, we have more than 140,000 images, and each of them has at least one transit event. Then, we randomly split the data into a training set and a validation set at a ratio of 85\% and 15\%. Different from the previously preserved test set, our training and validation sets are split on the image level to make the parameters of the transits evenly distributed in the training and validation sets as much as possible.

\subsection{Network Architecture}\label{subsec:network}
\begin{figure}
    \centering
    \includegraphics[scale=0.9]{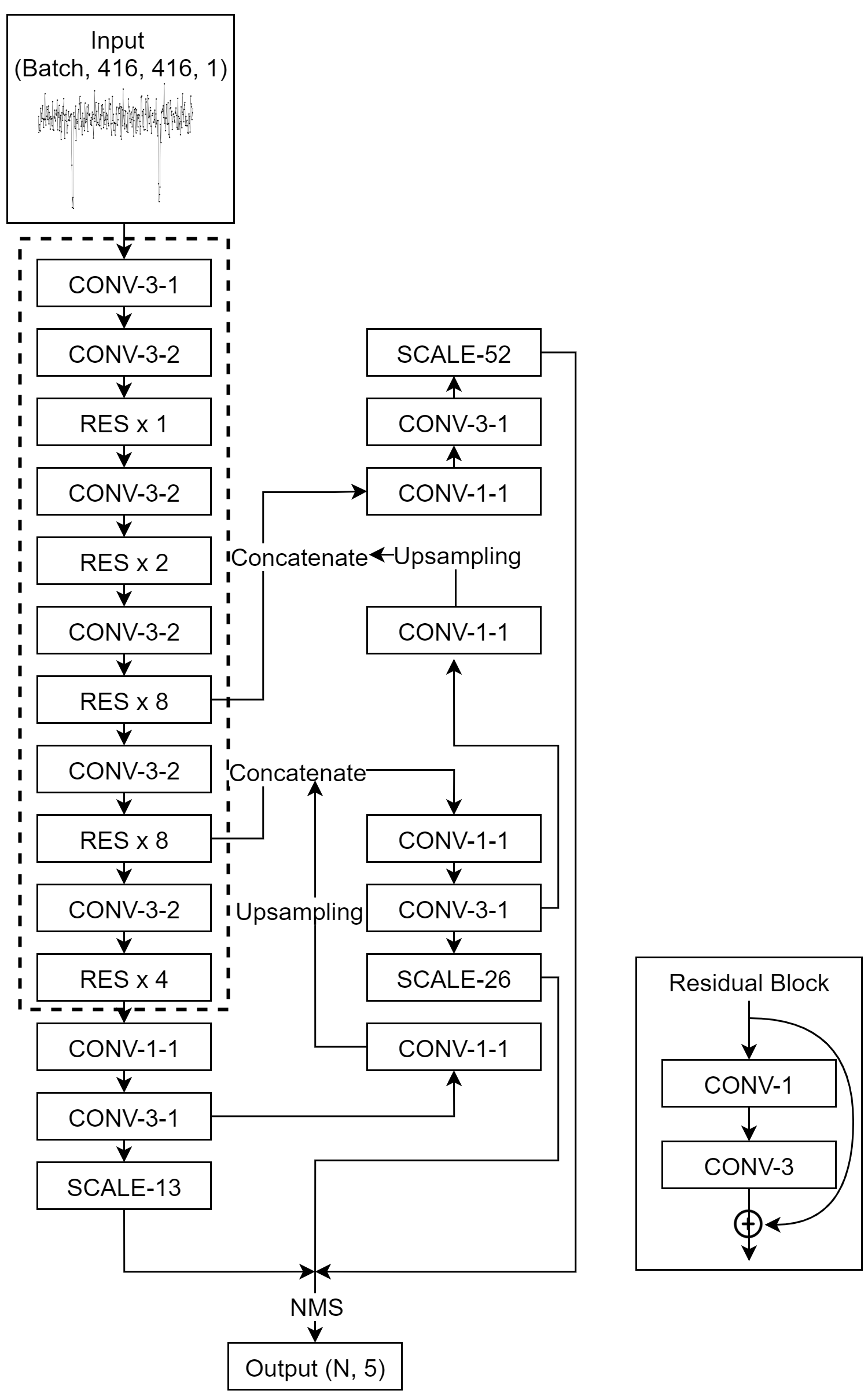}
    \caption{Network architecture. Our network structure is based on a standard YOLOv3 architecture with a slightly modified input channel and output dimension. The dashed line region indicates the Darknet-53 backbone. Each convolution block is shown as a CONV, it is combined by a 2D convolution layer, a batch normalization layer, and a Leaky ReLU function \citep{maas2013rectifier}. The numbers split by dashes of a convolution block show the size and stride of the convolution layer. Each residual block is abbreviated as RES. As shown on the right side of the figure, a residual block is combined by two convolution blocks with a shortcut. The scale block is shown as SCALE, and the following number is the size of each scale. Similar to the residual block, each scale block has two convolution blocks but without a shortcut. Combining the results of the three scales to calculate the NMS, the final output is the coordinates and confidence level of N-detected bounding boxes.}
    \label{fig:network-arch}
\end{figure}

We do not need to invent a new object detection network because there are already a lot of well-proven algorithms and tools in the computer vision field. We choose the well-established You Only Look Once, Version 3 architecture \citep[YOLOv3;][]{Redmon2018arXiv180402767R}, which has a wide range of applications \citep[e.g.,][]{choi2019gaussian,tian2019apple,Yurtsever9046805} and performs better than previous algorithms (e.g., R-CNN; \citealt{Girshick2014}, Faster R-CNN; \citealt{Ren2017}) in terms of small object recognition and computational speed. Our network architecture is shown in Figure \ref{fig:network-arch}. It has one input head and an output of three scales for large, medium, and small objects. The final output combines the results of three scales through the Non Maximum Suppression (NMS) algorithm, which is a standard way for only keeping best bounding boxes \citep{Rosenfeld1671883}. The main procedure of NMS is 
to iteratively preserve the boxes with the highest confidence score and then remove other boxes depending on a threshold.

Original YOLOv3 network is designed for detecting daily common objects, so they set their prior anchor boxes based on some competition data sets (e.g., PASCAL Visual Object Classes Challenge and Microsoft Common Objects in Context). 
Anchor boxes are crucial parameters in YOLO, they are a set of predefined bounding boxes of a certain height and width.
However, in our work, we are aiming at identifying transits, the size of anchors should be determined by the characteristics of the signals, and since we have only one class, too many anchor boxes are overkill for our detection. Thus, we measured the height and width distribution of all the boxes in our data set, and then we manually selected three anchors to roughly get close to the size of most of the transiting signals, while considering assigning different sizes of anchors for different scales. Finally we reset the width and height of anchor boxes to (0.1, 0.9), (0.05, 0.7), and (0.02, 0.3) for three scales accordingly. The width is significantly lower than the height, consistent with the basic pattern of a transit. We try adding more anchors but it does not improve the performance. We also compare the original YOLOv3 anchors with our modified version and find that the modified one is better in terms of transit identification.

Usually, the confidence score of a bounding box is trained to equal the intersection over union (IOU) between a predicted box and the ground truth. 
IOU is defined as a ratio of the areas of intersection and union between the predicted bounding box and the ground-truth bounding box.
IOU describes the accuracy of a predicted box; the higher the IOU, the more accurate the overlap with the target. 
However, the range of the target box is artificially selected by us, so the high overlap does not have much physical sense. Instead, we consider a more physical expression of the confidence score to be the S/N.
We can easily calculate the S/N for a single-transit event, and the S/N can be naturally converted to a confidence score with a normalization method. In this work, our confidence scores $C_\mathrm{S/N}$ of training data are calculated by $C_\mathrm{S/N} = 1 - \exp(-0.15 \ \mathrm{S/N})$. The $C_\mathrm{S/N} \approx 0.6$ when the S/N is 6, and then rapidly increase to 1, matching our naked eye's intuitive perception and previous research thresholds \citep[e.g.,][]{Kov2002A&A...391..369K, Kunimoto2018AJ....155...43K}. A different normalization method is also acceptable for different requirements.

Then, we modify the original YOLOv3 loss function because of the different number of anchor boxes and confidence scores. As shown in Equation \ref{eq:loss}, loss function of each scale is a combination of three parts and is defined as

\begin{eqnarray}\label{eq:loss}
L & = & L_\mathrm{coord} + L_\mathrm{obj} + L_\mathrm{noobj} \nonumber ~,\\ 
& = & \lambda_{\mathrm{coord}} \frac{1}{N} \sum_{i=0}^{S^2} \mathbbm{1}_{i}^{\mathrm{obj}}[(x_i-\hat{x}_i)^2 + (y_i-\hat{y}_i)^2 ] \nonumber \\
& + & \lambda_{\mathrm{coord}} \frac{1}{N} \sum_{i=0}^{S^2} \mathbbm{1}_{i}^{\mathrm{obj}}[(w_i-\hat{w}_i)^2 +(h_i-\hat{h}_i)^2 ]  \nonumber \\
& - & \frac{1}{N} \sum_{i=0}^{S^2} \mathbbm{1}_{i}^{\mathrm{obj}} [C_\mathrm{S/N} \log (C_i) + (1 - C_\mathrm{S/N}) \log (1 - C_i))] \nonumber \\
& - & \lambda_{\mathrm{noobj}} \frac{1}{N^{\prime}} \sum_{i=0}^{S^2} \mathbbm{1}_{i}^{\mathrm{noobj}} \log (1-C_i)~.
\end{eqnarray}

The sum of the first two terms is the coordinate mean squared error (squared L2 norm) loss. $x_i$, $y_i$, $w_i$, $h_i$ denote the predicted coordinates of boxes in the cell $i$, similarly, $\hat{x}_i$, $\hat{y}_i$, $\hat{w}_i$, $\hat{h}_i$ denote the target coordinates of boxes. $S$ is the scale size. $\mathbbm{1}_{i}^{\mathrm{obj}}$ equals to 1 when there is an object in the cell $i$, otherwise 0; $\mathbbm{1}_{i}^{\mathrm{noobj}}$ equals 1 when there is no object in the cell $i$. $N$ is the number of object cells for calculating the mean value. The third and fourth terms show the binary cross-entropy losses of the object and no object cells.
Binary cross-entropy is a commonly used cross-entropy loss for binary classification applications.
$C_i$ is the predicted confidence score of whether there is an object or not. 
Similar to $N$, $N'$ is the number of cells without object. $\lambda_{\mathrm{coord}}$ and $\lambda_{\mathrm{noobj}}$ are the constants to increase the attention of certain parts. In our work, we choose $\lambda_{\mathrm{coord}} = \lambda_{\mathrm{noobj}} = 10$ to ensure the balance of those three losses. Compared with original YOLOv3 loss function, we do not have the class loss, because we only focus on transit signals.

\subsection{Training Steps}\label{subsec:training}
There are two key metrics precision and recall. Precision is defined as $\mathrm{TP/(TP+FP)}$, and recall is defined as $\mathrm{TP/(TP+FN)}$. The TP, FP, and FN indicate true positive, false positive, and false negative. Average precision (AP) is calculated by the area under the precision-recall (PR) curve.
$\mathrm{AP}_{50}$ means AP with the IOU threshold greater than 0.5, $\mathrm{AP}_{70}$ means AP with the IOU threshold greater than 0.7, and so does $\mathrm{AP}_{90}$.
We choose the mean value of $\mathrm{AP}_{50}$, $\mathrm{AP}_{70}$,  and $\mathrm{AP}_{90}$ as our training evaluation metric ($\mathrm{AP}_{\mathrm{mean}}$). The $\mathrm{AP}_{50}$ is the baseline of many algorithms because IOU larger than 0.5 is sufficient for most ordinary target detection tasks. We consider $\mathrm{AP}_{70}$ and $\mathrm{AP}_{90}$ because we want the parameters of the box to provide better help for subsequent applications.

After several trials based on the performance on the validation set, we choose the Adam optimizer to optimize the total loss $L$. The initial learning rate is $10^{-4}$, and the weight decay is $10^{-4}$. A lower learning rate would increase the training steps significantly. We also reduce the learning rate by a factor of 2 when our metric has stopped improving for two epochs. The epoch is the number of passes of the entire training data set. In our work, each epoch has 1265 iteration steps.
To improve training efficiency, we input a batch of training examples in each iteration step. Our batch size is 96, lower batch size performs worse. The NMS threshold is 0.1, the IOU threshold is 0.5, and the confidence score threshold is 0.6.
Each epoch takes about 20 minutes on an NVIDIA Tesla V100 GPU.

\begin{figure}
    \centering
    \plotone{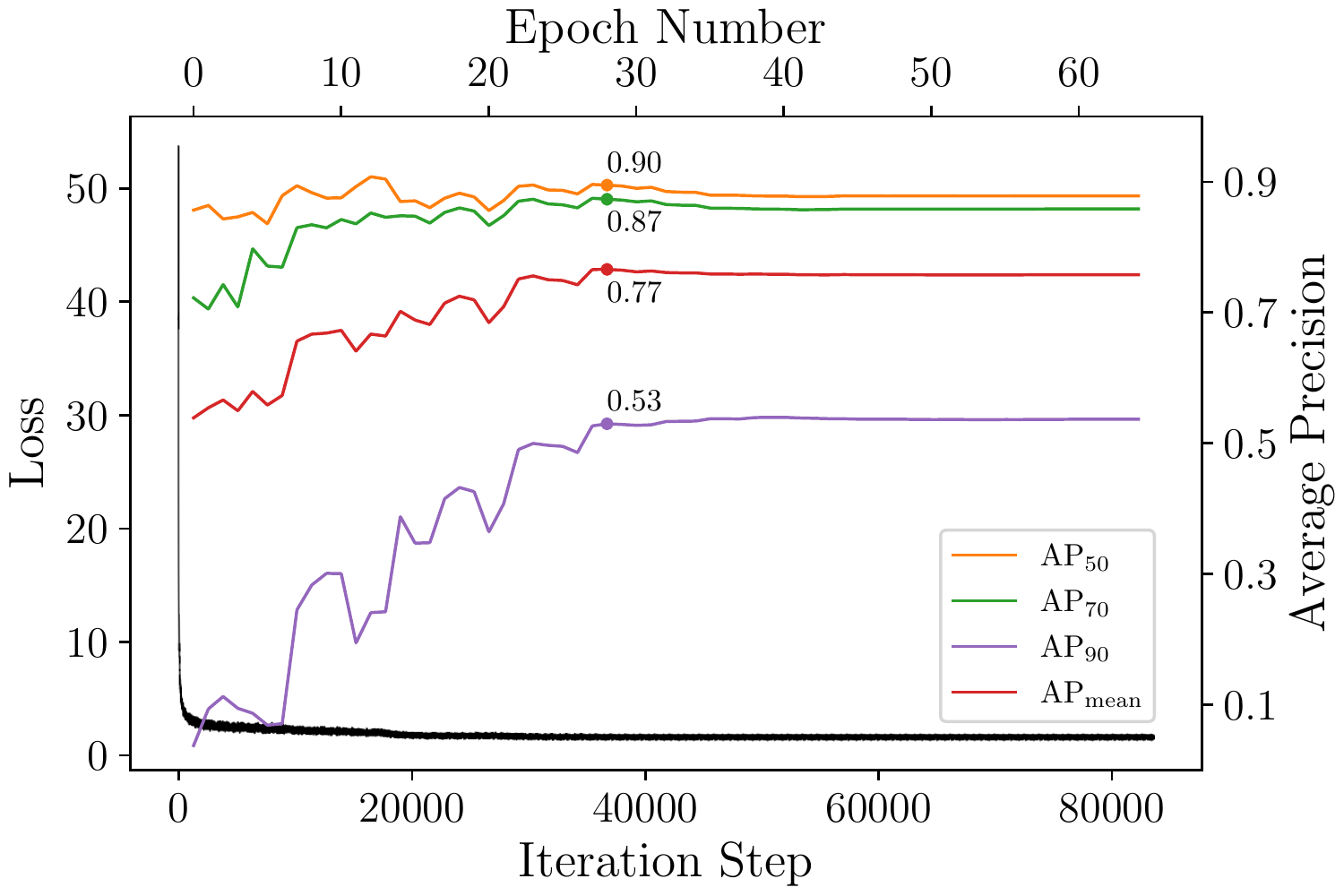}
    \caption{Training steps on the Kepler light curves. The black line is the loss and the APs are plotted as different colors for different IOU thresholds. The AP values of the best model are marked and written on the figure.}
    \label{fig:kepler-training}
\end{figure}

\begin{figure}
    \centering
    \plotone{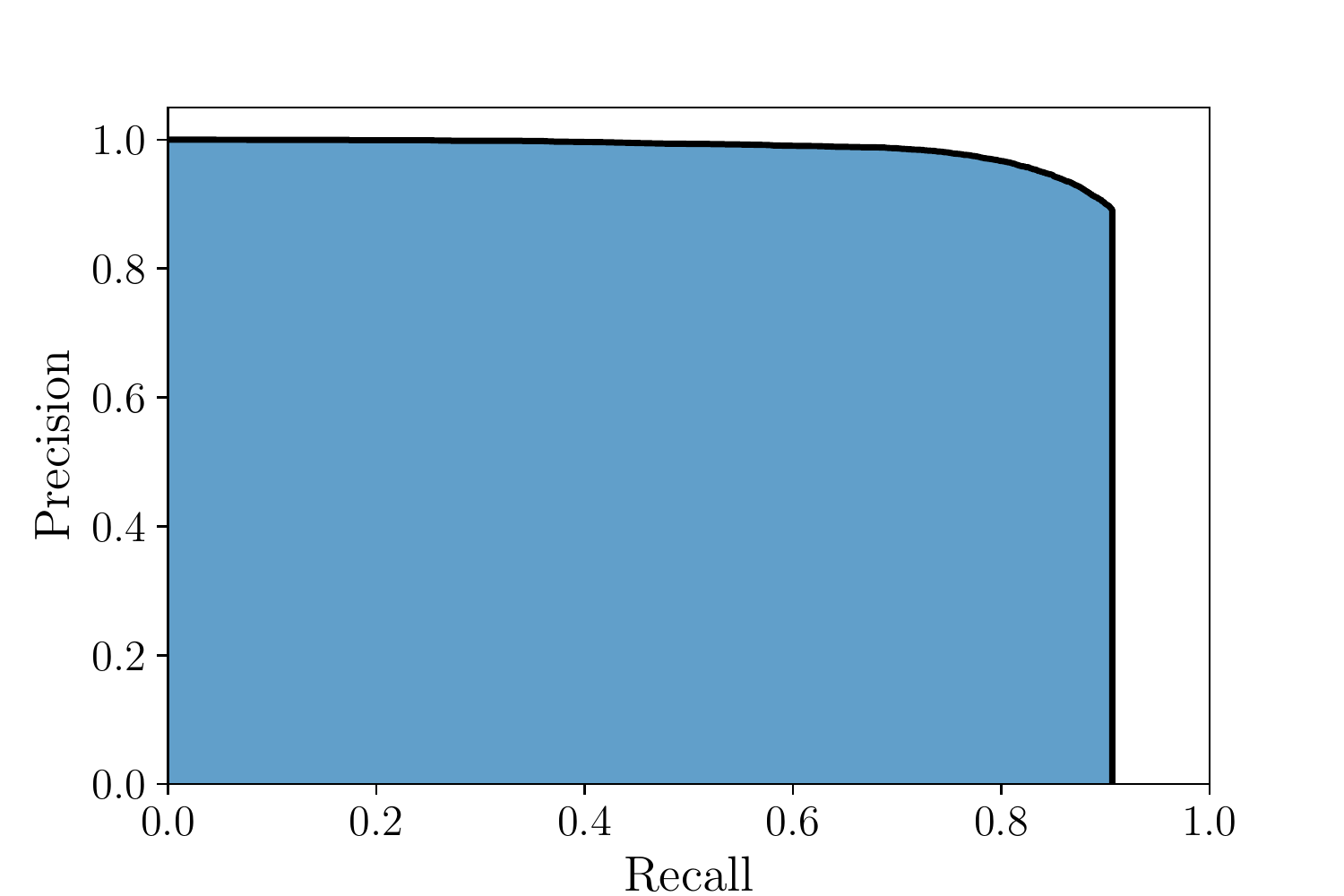}
    \caption{The precision-recall curve for IOU $>$ 0.5 of the best model. The blue shaded area is the $\mathrm{AP}_{50}$. The sharp drop at 0.9 is due to our relatively high confidence threshold (0.6), making recall only go up to 0.9.}
    \label{fig:pr-curve}
\end{figure}

Figure \ref{fig:kepler-training} shows the losses, $\mathrm{AP}_{50}$, $\mathrm{AP}_{70}$, $\mathrm{AP}_{90}$ and $\mathrm{AP}_{\mathrm{mean}}$, vary during the training. Those APs are calculated on the validation data, and we can see them increase and then keep constantly after longer epochs. Due to overfitting, the loss on the training set continues to drop, but the metrics on the validation set are no longer rising. Therefore, we choose the model with the largest $\mathrm{AP}_{\mathrm{mean}}$ on the validation set as the best model. The PR curve of the best model is shown in Figure \ref{fig:pr-curve}. Our final best model has an $\mathrm{AP}_{50}=0.9$, which is an excellent performance for an object detection task. The best model file saves the states of the network trained from Kepler data, which can be directly downloaded from \href{https://registry.china-vo.org/resource/101079}{DOI: 10.12149/101079}.

\subsection{Model Tests and Comparisons}\label{subsec:kepler-test}
\begin{figure}
    \centering
    \plotone{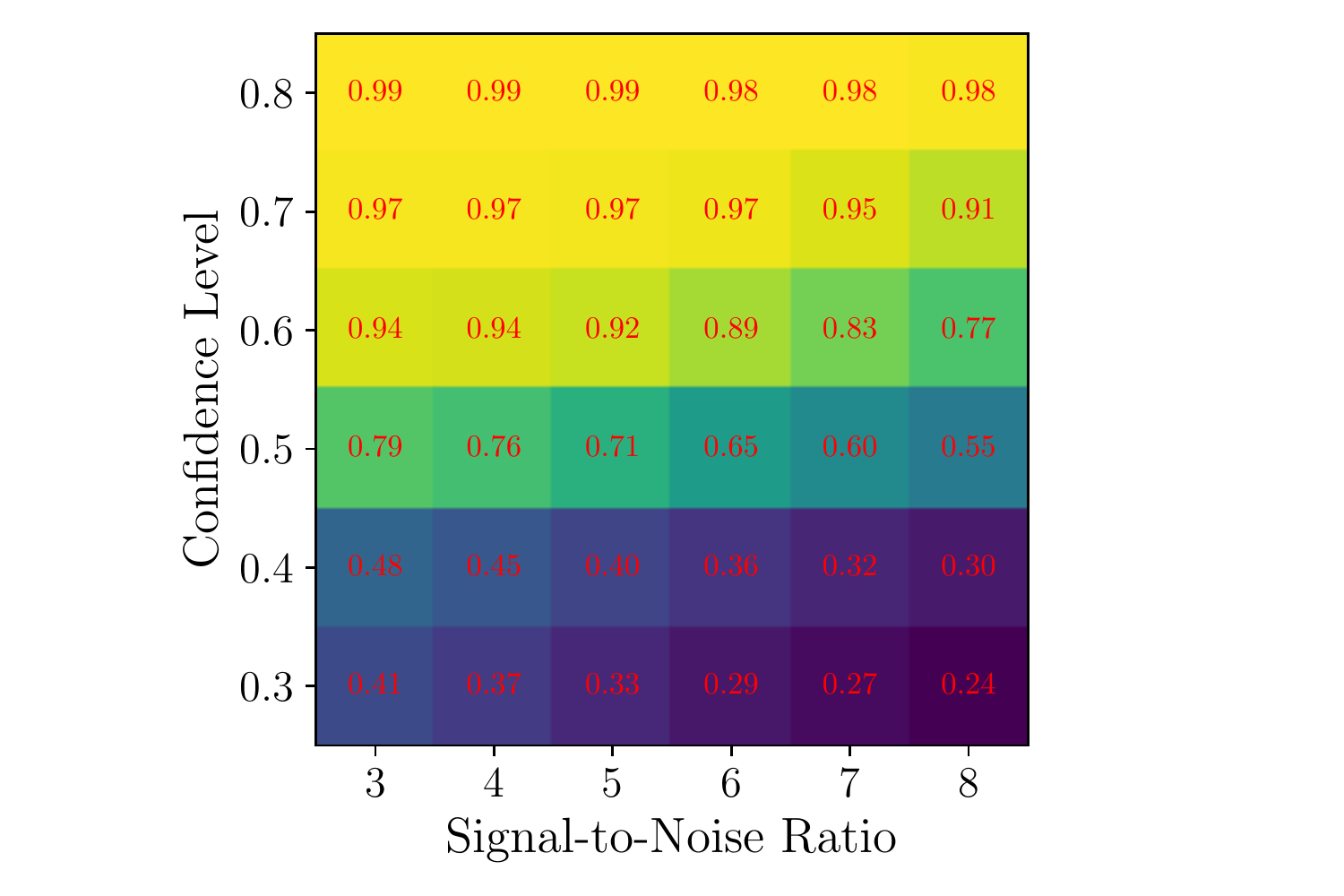}
    \caption{Precision matrix of the best model trained from Kepler on the test set.}
    \label{fig:kepler-precision}
\end{figure}

\begin{figure}
    \centering
    \plotone{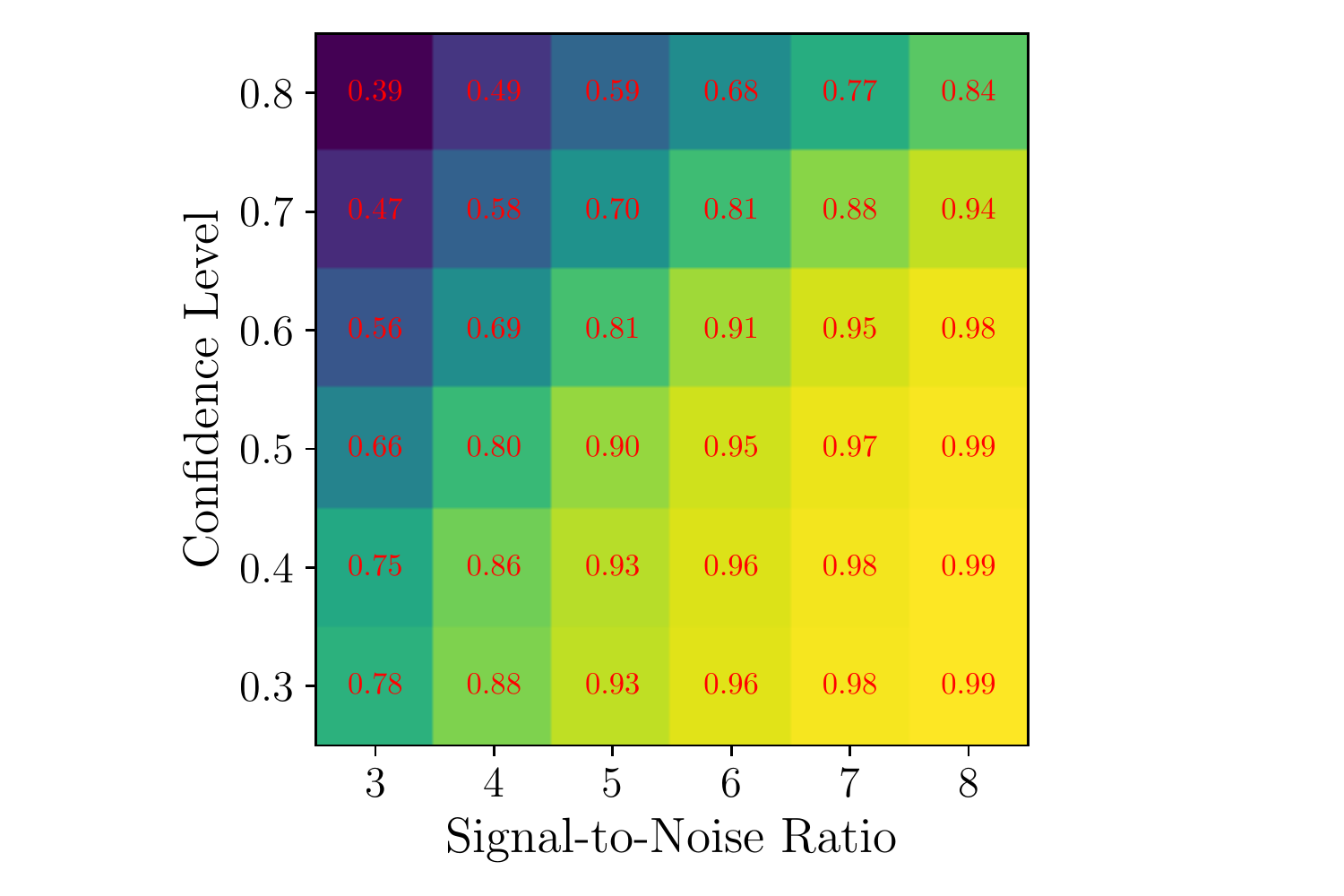}
    \caption{Recall matrix of the best model trained from Kepler on the test set.}
    \label{fig:kepler-recall}
\end{figure}

After finishing the training, we need to test the performance of the model. Although our high $\mathrm{AP}_{\mathrm{mean}}$ on the validation set gives us confidence, we still want to know its performance on real detection when facing transits with different S/Ns.
To do so, we apply our best model on 100 previously reserved exoplanet hosts. Consistent with the real detection task, we use the detection function implemented in our {\tt Deep-Transit} package with default parameters. To use more physical criteria, we treat a detected box as a true positive when the time of the transiting midpoint locates in the box range. We apply our test on different confidence levels and different S/Ns of transits. Figure \ref{fig:kepler-precision} and Figure \ref{fig:kepler-recall} show the precision and recall matrices accordingly. These two matrices present a comprehensive evaluation of the detection ability of our model. We can also easily recognize that the trade-off between precision and recall. For the detected confidence scores higher than 0.6, our model's precisions can achieve high to 0.9 in most cases. Higher confidence scores allow precisions close to 1. For the confidence scores lower than 0.5, the recalls are higher than 0.8 in most cases. Lower confidence scores also make recalls close to 1 for S/Ns higher than 6. Therefore, our model has high practicality and can control the precision and recall of detection by adjusting the confidence threshold.
To have a balance between precision and recall, 0.6 can be a Goldilocks score in most cases. To perform an extensive search, 0.5 is a practical choice.

We also compare our model with two practically used baselines. Unlike other commonly used exoplanet detection algorithms (periodogram-based algorithms such as box-fitting least squares), our model lacks period information. Therefore, a more appropriate baseline is the algorithm of single-transit detection. 
One of the most straightforward and conventional single-transit detection algorithms is a transit fitting with a moving window, which is applied by most single-transit event searches of Kepler and K2 with only minor differences \citep[e.g.,][]{Foreman-Mackey2016AJ....152..206F,Osborn2016MNRAS.457.2273O,Kawahara2019AJ....157..218K}. We implement this algorithm by following \citet{Osborn2016MNRAS.457.2273O}. 
Our transit model is an analytic model with quadratic limb darkening \citep[][]{Mandel2002ApJ...580L.171M}, and it is generated with the {\tt Batman}, which is an open-source Python package for fast calculation of exoplanet transit light curves. The window sizes and transit durations are exactly adopted from \citet{Osborn2016MNRAS.457.2273O}. We applied this algorithm to our test set. For transits with S/N $>$ 6, the precision is 0.59 when the recall reaches 0.91 by tuning the reduced $\chi^2$ threshold. The higher number of false positives is also confirmed in \citet{Kawahara2019AJ....157..218K}, this is also why single-transit detection usually requires visual inspections to rule out false positives. Our network is based on computer vision techniques, thus we have fewer false positives.

Another baseline is given by the naked eyes. Many exoplanet detection pipelines require visual vetting to remove false positives \citep[e.g.,][]{Batalha2013ApJS..204...24B,Huang2013MNRAS.429.2001H,Guerrero2021ApJS..254...39G}. Also, as one of the most famous citizen science projects, Planet Hunters \footnote{\url{https://www.zooniverse.org/projects/nora-dot-eisner/planet-hunters-tess}} attracts volunteers to visually detect exoplanets. According to their papers \citep[][]{fischer2012MNRAS.419.2900F,Eisner2021MNRAS.501.4669E}, the practical recovery rate of Planet Hunters is about 67\% for Kepler data, and 53\% for TESS objects of interest (TOI) data. The spurious transit boxes rate is about 10\%. Therefore, compared with the Planet Hunters, our network performs similar precision but higher recall. The result is reasonable because our trained model is more sensitive to transit signals compared to amateur volunteers.

\section{Transfer Learning to TESS}\label{sec:transfer}
As an ongoing survey, TESS provides higher duty cycle light curves than Kepler. Therefore, we also wish to apply our well-trained model to the TESS data.
Usually, deep learning algorithms can only achieve high performance on a specific task. It means that our model trained on Kepler can not be directly applied to TESS. However, since the main difference between TESS and Kepler is their photometric accuracy and sampling rates, we can create a well-performing model for TESS data based on the pretrained Kepler model. To do so, a transfer learning technique can be helpful.

Transfer learning inherits existing relevant data or models while building a new machine-learning model. In practice, transfer learning is a standard way when there are not enough training data, or the source and target domains have some similarities but are not identical \citep{pan2009survey}. Therefore, when considering the short observation time of TESS, resulting in a small amount of data, transfer learning is an appropriate choice. To achieve this, we need to prepare the data set for TESS and then load the model already trained on the Kepler data as a pretrained model. The training on TESS data only requires a fine-tuning of the pretrained model.

\subsection{Training and Evaluation}\label{subsec:transfer-train}
\begin{figure}
    \centering
    \plottwo{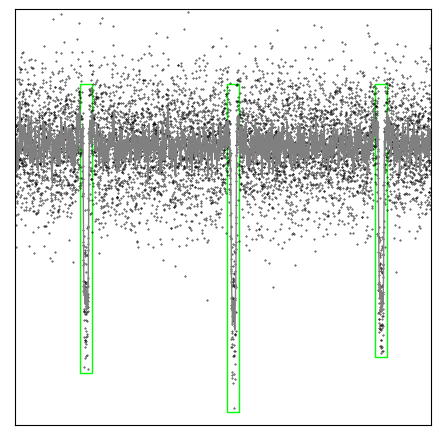}{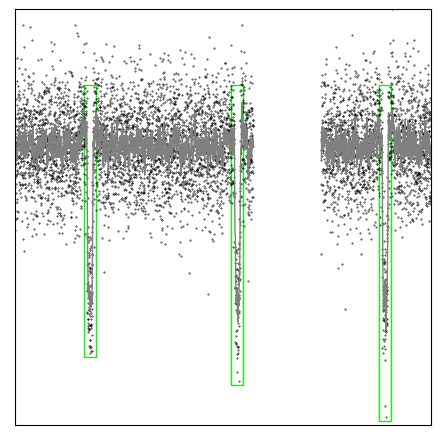}
    \caption{Two examples of TESS training and validation data. The bounding boxes are shown as green boxes. Note that the actual sample does not have a border.}
    \label{fig:tess-sample}
\end{figure}

First, we create the TESS data set similarly as we create the Kepler data set (described in Section \ref{subsec:data-prep}). Since there are only over 100 confirmed exoplanets detected by TESS, our training and validation data set are created from the TOI catalog \citep[][]{Guerrero2021ApJS..254...39G}. The main difference in the data set preparation is that the TESS images are plotted using scatter points with a 30 day moving average gray line overlaid. The points are 1 pixel points and the line width is 1 point. Figure \ref{fig:tess-sample} shows two examples of the TESS training data set.

\begin{figure}
    \centering
    \plotone{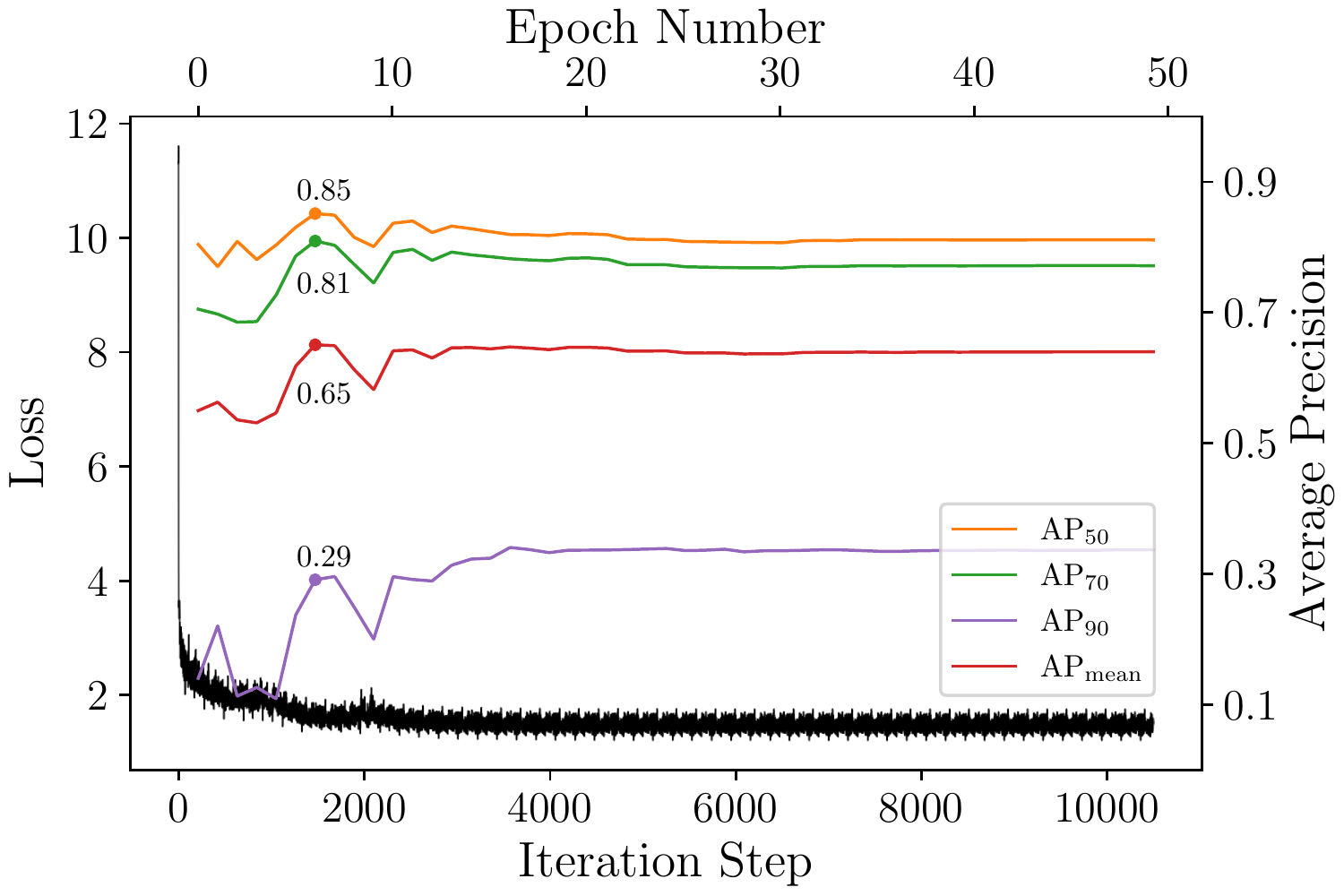}
    \caption{Training steps of transfer learning on TESS data. Same as Figure \ref{fig:kepler-training}.}
    \label{fig:train-transfer-tess}
\end{figure}

Then, we use the same network structure and load our pretrained Kepler model to initialize the network without freezing any layers. As the model converges quicker, we reduce the learning rate by a factor of 2 when our metric has stopped improving for one epoch. The performance during training steps is shown in Figure \ref{fig:train-transfer-tess}.
Similarly, as described in Section \ref{subsec:training}, we choose epoch 6 as the best performance model. Also, the best model saves the states of the network trained from TESS data, which can be directly downloaded from \href{https://registry.china-vo.org/resource/101079}{DOI: 10.12149/101079}.

\begin{figure}
    \centering
    \plotone{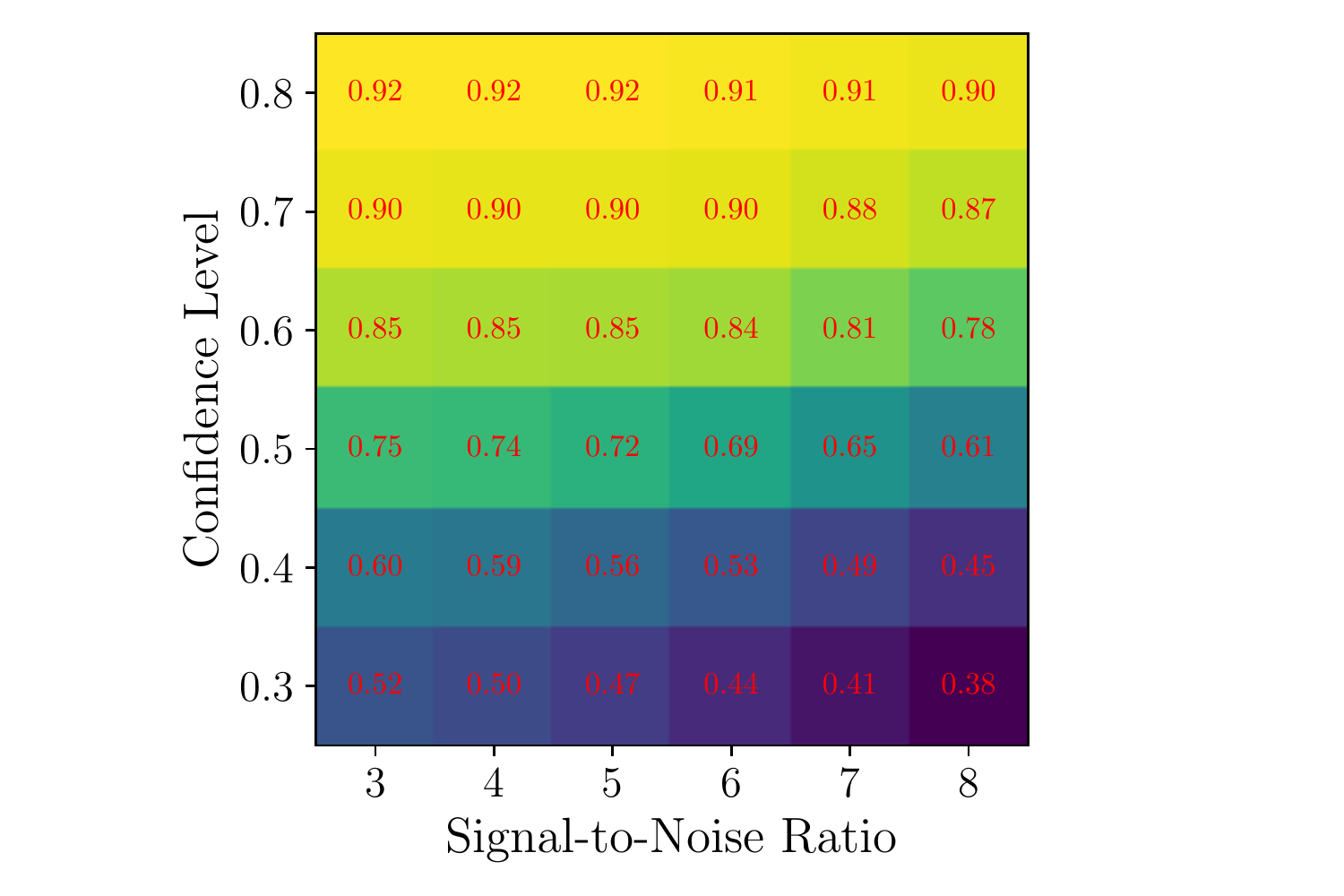}
    \caption{Precision matrix of the best model trained from TESS on the test set.}
    \label{fig:tess-precision}
\end{figure}

\begin{figure}
    \centering
    \plotone{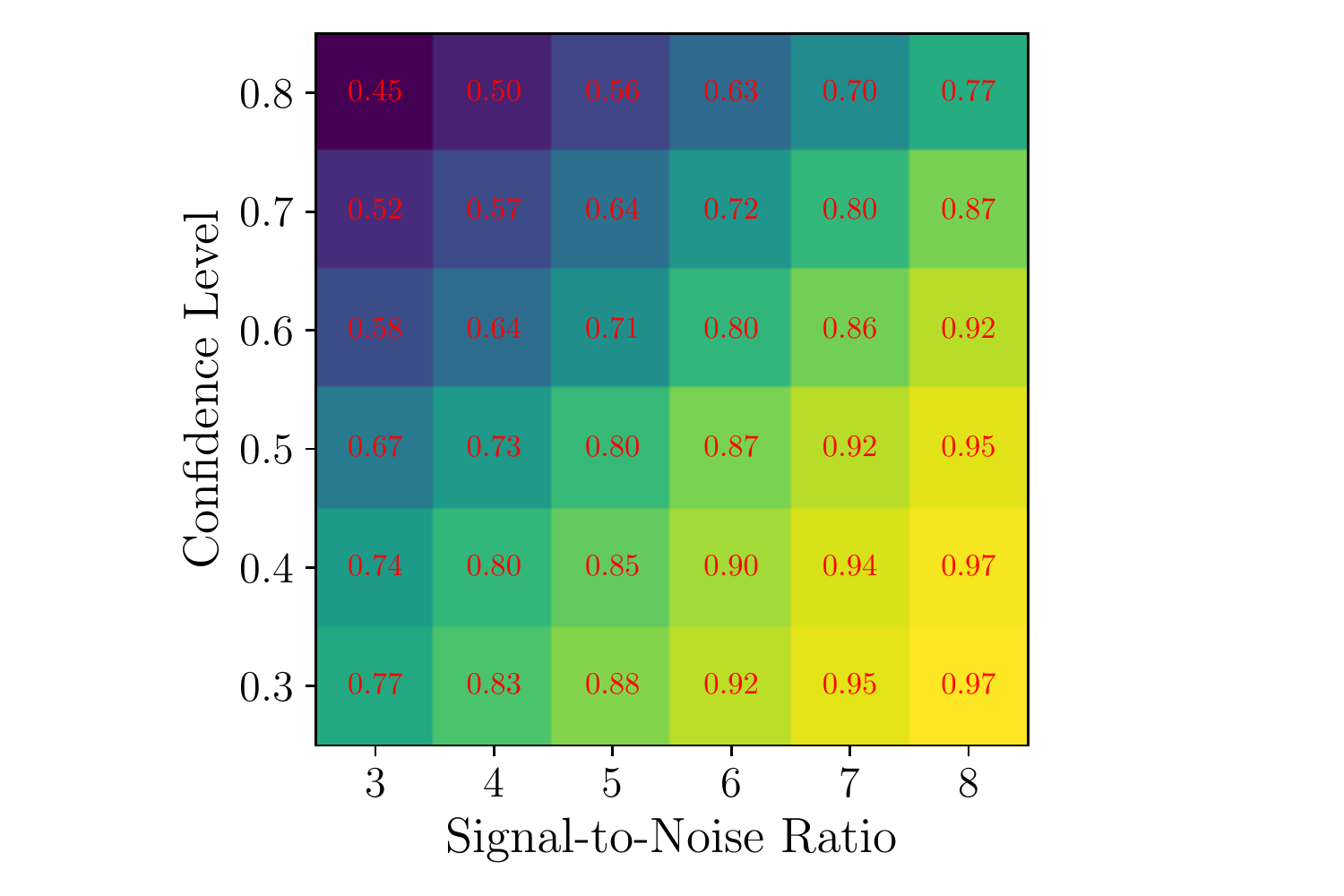}
    \caption{Recall matrix of the best model trained from TESS on the test set.}
    \label{fig:tess-recall}
\end{figure}

Finally, we test the performance of our model using light curves of confirmed TESS exoplanet hosts. Due to the relatively small amount of data in TESS, we manually adjust some parameters of the detrending method. Like with the previous Kepler model, the test results are shown as two matrices (see Figure \ref{fig:tess-precision} and Figure \ref{fig:tess-recall}). Their general performance is good but relatively lower than Kepler, which might be caused by the lower data quality of TESS. The recall is better than precision, makes it more suitable for a large-scale transit search.

\section{Application}\label{sec:application}
After the training and evaluation of the model is completed, our algorithm can have some good practical applications. 

Once the bounding boxes with confidence scores of transits are successfully detected, the width and height of boxes can be used as a rough estimation for the duration and depth of the transits.
For example, Kepler-297 (KIC 11122894) has three transit candidates and two of them are confirmed \citep{Lissauer2011ApJS..197....8L,Rowe2014ApJ...784...45R}.
Table \ref{tab:kepler-297} list some of their parameters from \citet{Rowe2014ApJ...784...45R}.
\begin{deluxetable}{cccc}
\label{tab:kepler-297}



\tablecaption{Table of parameters of transits of Kepler-297 from \citet{Rowe2014ApJ...784...45R}}


\tablehead{\colhead{Parameters} & \colhead{} & \colhead{} & \colhead{}} 

\startdata
KOI &  1426.01 &  1426.02 &  1426.03 \\
Kepler ID &   Kepler-297 b &  Kepler-297 c &   \\
Period (days) &  38.871826 &   74.920137 &  150.018303 \\
$R_\mathrm{p}/R_*$ &   0.02876 &   0.06543 &  0.59798 \\
Transit Depth &  940.4 &  4150.1 & 4437.6\\
Transit Duration &  6.125 & 4.794 & 4.400
\enddata




\end{deluxetable}

\begin{figure*}
    \centering
    \includegraphics[width=\textwidth]{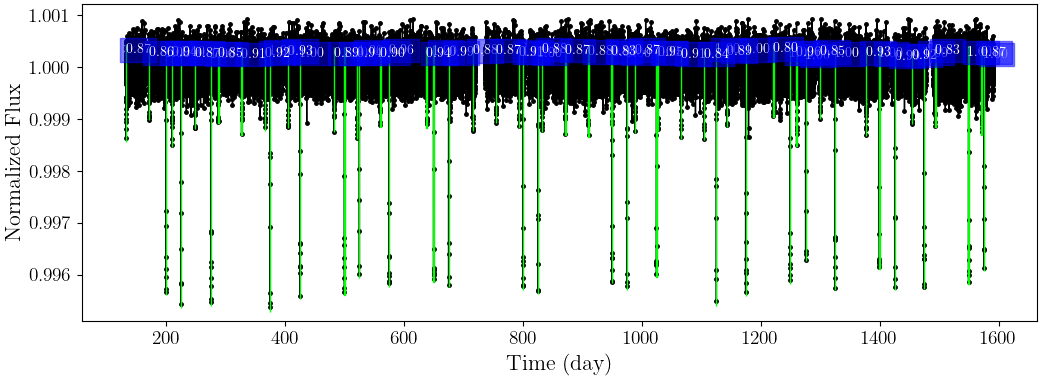}
    \caption{Light-curve and transit detection results of Kepler-297. The green boxes are the detected bounding boxes, and their top square blue labels indicate the confidence scores.}
    \label{fig:multi-lc}
\end{figure*}

\begin{figure}
    \centering
    \includegraphics[width=0.8\textwidth]{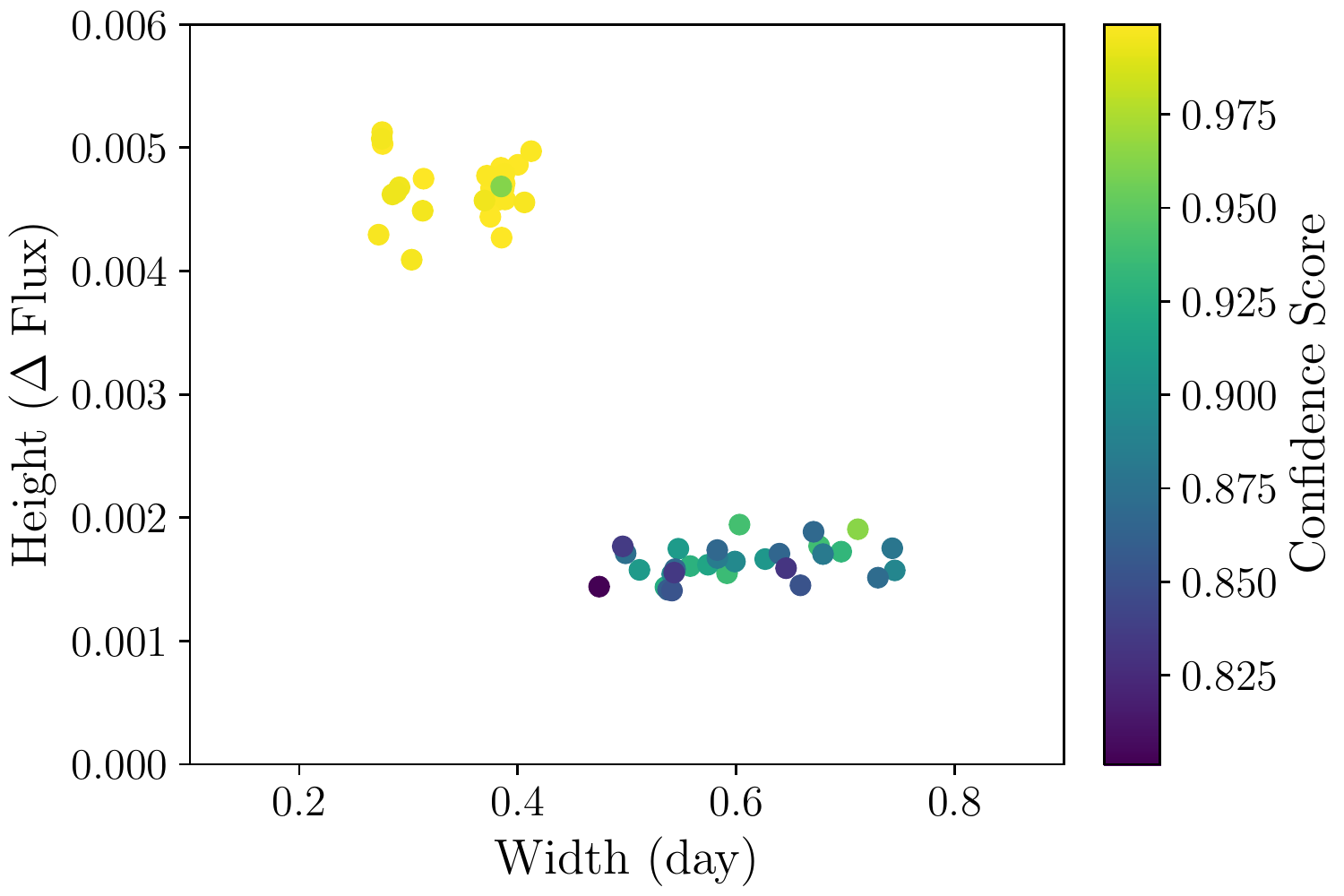}
    \caption{Scatter plot of the width and height of the detected boxes. The color of the points indicate the confidence level.}
    \label{fig:cluster}
\end{figure}

\begin{figure}
\gridline{\fig{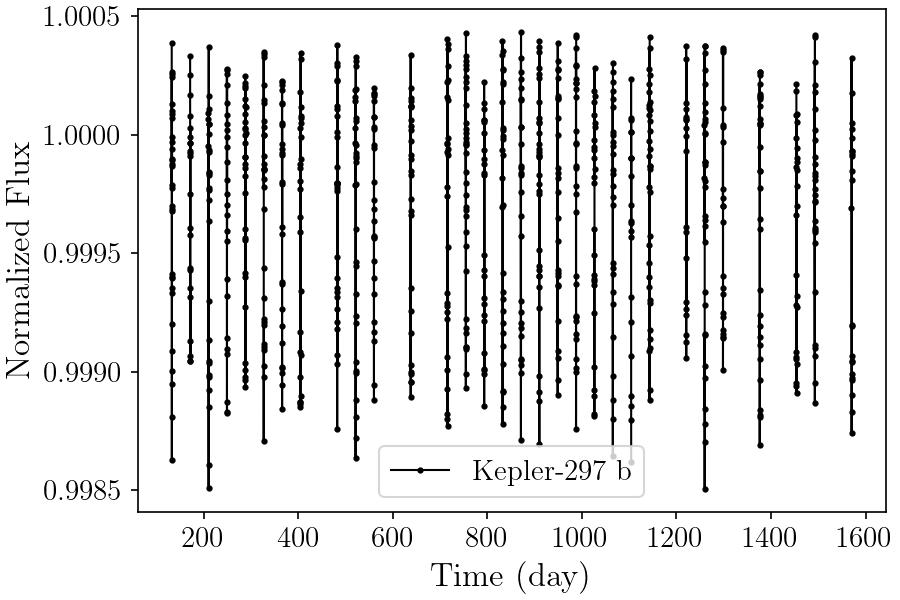}{0.6\textwidth}{(a)}}
\gridline{\fig{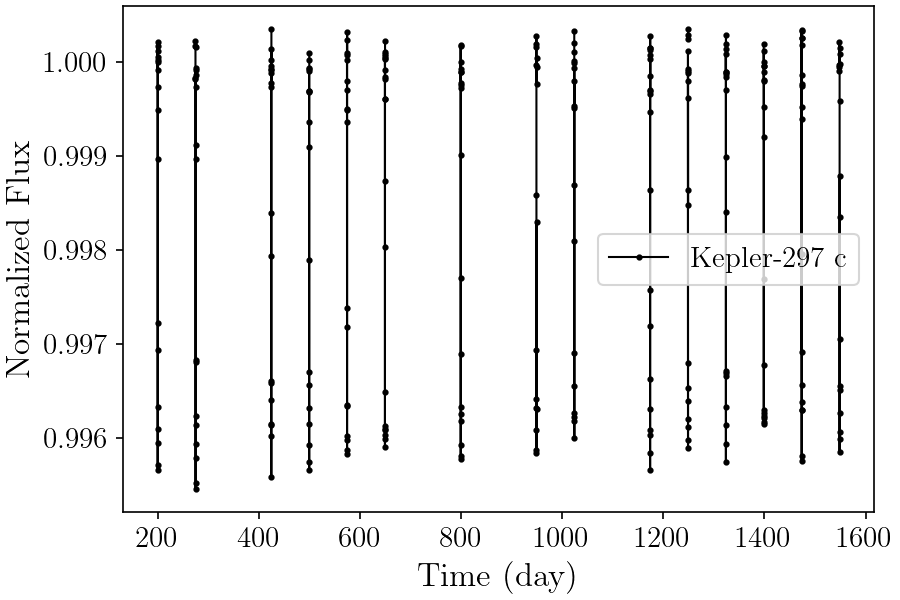}{0.6\textwidth}{(b)}}
\gridline{\fig{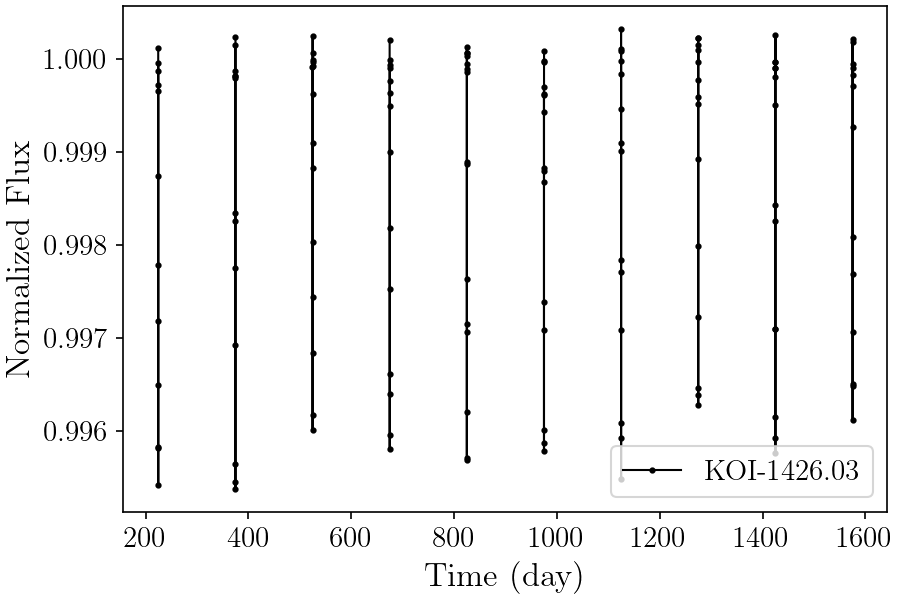}{0.6\textwidth}{(c)}}
\caption{Selected light curves from our detected bounding boxes. (a) is the selection of detected transiting events of Kepler297b. (b) is the selection of detected transiting events of Kepler297c. (c) is the selection of detected transiting events of KOI-1426.03.}
\label{fig:three-clusters}
\end{figure}

We can apply our detection method to the 4 yr light curve of Kepler-297 to reproduce its multiplicity. Figure \ref{fig:multi-lc} shows the flattened light curve and our detected bounding boxes. The confidence threshold is 0.8. After we put the width and height of the detected boxes into a scatter plot, we can see two obvious clusters in Figure \ref{fig:cluster}. If we look closely, the yellow points in the upper-left corner also seem can be divided into two parts. Actually, they have similar depth and width but different periods. 
By selecting the light curve in the detected bounding boxes, we can perform a more focused period detection. As shown in Figure \ref{fig:three-clusters}, the periods of those three clusters can be estimated from the middle time directly.

\begin{figure*}
    \centering
    \gridline{\fig{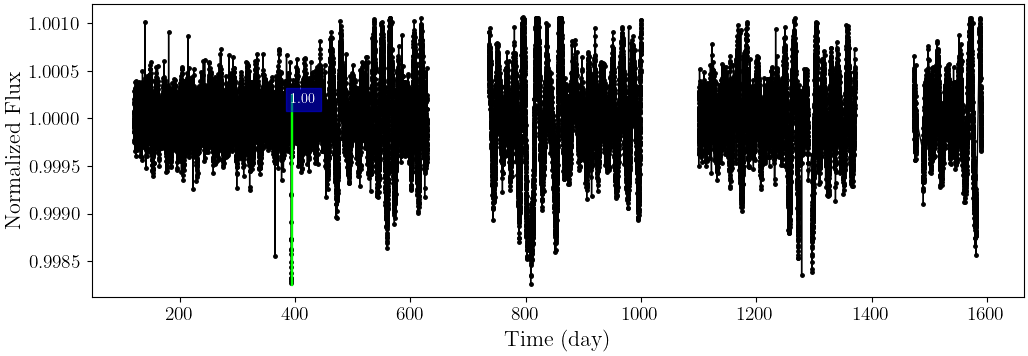}{\textwidth}{(a)}}
    \gridline{\fig{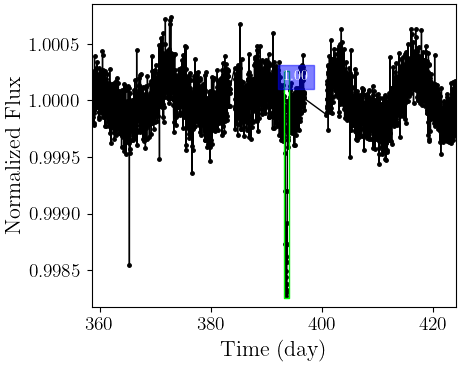}{0.5\textwidth}{(b)}}
    \caption{Light curve of KOI-1174. (a) is the 4 yr light curve of KOI-1174. (b) is the zoom-in of the detected region. The green box is the detected bounding box and the confidence score is written in the semitransparent blue box.}
    \label{fig:single-transit}
\end{figure*}

Since our algorithm is complete and accurate to a single transit with high S/N $>$ 7, it has a strong potential for single-transits detection. 
\citet{Kawahara2019AJ....157..218K} yields 67 single-transit candidates in Kepler light curves from previous work \citep[e.g.,][]{Uehara2016ApJ...822....2U,Foreman-Mackey2016AJ....152..206F}.
Nearly all of them can be detected by our model with proper detrending (the default window size 0.5 day will remove some long duration transits). Figure \ref{fig:single-transit} shows an example of KOI-1174, which has a single-transiting exoplanet candidate. Our detection algorithm can easily find the only transit with a confidence score higher than 0.8.
Finding single transits becomes even more important due to the observation mode of TESS \citep[e.g.,][]{Cooke2018A&A...619A.175C,Villanueva2019AJ....157...84V,Yao2021AJ....161..124Y}. Our model can be a powerful tool for future single-transits detection work.

\section{Discussion}\label{sec:discuss}
From the human perception, our method is intuitive and can help some visual searches and inspections. For example, the individual vetting and group vetting of TOIs could apply methods similar to ours to reduce manual effort and increase efficiency. Also, some citizen science projects like Planet Hunters and Citizen ASAS-SN \footnote{\url{https://www.zooniverse.org/projects/tharinduj/citizen-asas-sn/classify}} can be partly replaced by our method. An alternative approach is to use the results of citizen projects as the training set for the model. 

From the view of data analysis, our methods of converting 1D time series into 2D images can be seen as a kind of feature engineering, which is widely used in the field of signal processing. For example, the discrete cosine transform converts data from time domain to frequency domain and it is commonly used in many traditional and deep learning areas, such as computer vision \citep[e.g.,][]{Zhang8960425}, voice recognition \citep{Bae7603830} and electroencephalogram signal detection \citep{rundo2019innovative}. Our approach implicitly includes some prior and feature enhancements from our own visual perception. This provides the direction for the optimization of our algorithm: clearer and more intuitive illustration or presentation of data for humans can yield better results on the computer.

Considering that the observation windows of ground-based telescopes are often affected by weather, season, etc., the sampling interval is more complicated. Our method can be seamlessly extended to the ground-based photometric survey, like ASAS-SN, ZTF, Vera C. Rubin Observatory, etc.  In particular, multiband photometry can be naturally implemented as the multiple channels of an image, providing richer information.

However, there are three known limitations in our current algorithm.
\begin{enumerate}
    \item Periodicity determination. Estimate period directly from the neural network is a quite difficult problem. Traditional algorithms like the box-fitting least squares are still more robust.
    \item The method of the detrending light curve has a significant impact on the S/Ns of transits. Though Tukey's biweight algorithm performs fairly robust in blind searches, its parameters still need to be adjusted for different types of trends and systematics. 
    \item Technically, our algorithm only detects transit-like signals, we do not introduce other possible false positives (e.g., eclipsing binaries) in our data set. Considering that many works are using deep learning algorithms to do the transit inspection (e.g., Astronet of \citealt{Shallue2018AJ....155...94S}; Exonet of \citealt{Ansdell2018ApJ...869L...7A}), our results can be the inputs of their networks to classify signals. 
\end{enumerate}

We have plans to solve the above difficulties in our future work. The lack of period information, intrinsic and external noises of the light curve, mainly limit the performance of our model for low S/N objects. An alternative solution is to simultaneously include the result of the box least squares periodogram and corresponding folded light curve. Some time-series modeling methods such as autoregressive moving average and its extensions \citep[e.g., autoregressive integrated moving average, autoregressive fractionally integrated moving average;][]{Carter2009ApJ...704...51C, Feigelson2018FrP.....6...80F,Caceres2019AJ....158...58C} can be helpful for removing stellar variations and systematics.

We will also extend our detection target to more classes of light-curve signals (e.g., rotation, pulsation, and eclipsing binary). By modeling those various types of detected objects, detrending algorithms can be more accurate, and false positives can also be identified. In addition, the data within the output box of the network can be used as the start of further detection and inspection of the signal, resulting in a complete imitation of human operation. However, some blended eclipsing binaries can be hardly discriminated from planetary transits only based on light curves. Usually, these contaminations are removed with some pixel-level inspection. Thus, we are trying to add some pixel vetting modules in our network, and they can be naturally applied in our 2D network.
With the development of the computer vision field, more advanced algorithms and structures (e.g., YOLOv5 and Transformer) can be a promising choice in our future work.

\begin{acknowledgments}
We appreciate the suggestions and comments of referee and editors that make this work more solid. We thank Masataka Aizawa for the helpful suggestions. We acknowledge
support from the National Key Research and Development Program of China (NKRDPC) under grant numbers 2019YFA0405000, B-type Strategic Priority Program of the Chinese Academy of Sciences, Grant No. XDB41000000, and from the National Science Foundation of China (NSFC, Nos. 1193000332, 1197030174).
Data resources are supported by China National Astronomical Data Center (NADC) and Chinese Virtual Observatory (China-VO). The authors acknowledge the Beijing Super Cloud Computing Center (BSCC) and the Tsung-Dao Lee Institute for providing GPU resources that have contributed to the research results reported within this paper. This paper includes data collected by the Kepler and TESS mission and obtained from the MAST data archive at the Space Telescope Science Institute (STScI). Funding for the Kepler and TESS mission is provided by the NASA Science Mission Directorate. STScI is operated by the Association of Universities for Research in Astronomy, Inc., under NASA contract NAS 5–26555.

\software{
Numpy \citep{2020NumPy-Array},
Matplotlib \citep{Hunter:2007},
PyTorch \citep{NEURIPS2019_9015},
Astropy \citep{Astropy2018AJ....156..123A,Astropy2013A&A...558A..33A}, 
Lightkurve \citep{Lightkurve2018ascl.soft12013L},
Wōtan \citep{Hippke2019A&A...623A..39H},
MegEngine \citep{megengine},
Jupyter Notebook \citep{Kluyver2016jupyter},
Batman \citep{Kreidberg2015PASP..127.1161K}
}
\end{acknowledgments}



\bibliography{deep-transit}
\bibliographystyle{aasjournal}


\listofchanges
\end{document}